\documentclass[a4paper]{amsart}
\usepackage{a4wide}
\usepackage[section]{placeins}
\usepackage{microtype}
\usepackage{amsmath}
\usepackage{amsthm}
\usepackage{amsfonts}
\usepackage{amssymb}
\usepackage[hidelinks]{hyperref}
\usepackage{graphicx}
\usepackage{todonotes}
\usepackage{enumitem}
\usepackage{array}

\title[Competition brings out the best]{Competition brings out the best: Modeling the frustration between curvature energy and chain stretching energy of lyotropic liquid crystals in bicontinuous cubic phases}

\author{Hao Chen}
\address{Freie Universit\"at Berlin, Institut f\"ur Mathematik, Arnimallee 2, 14195 Berlin, Germany}
\email{hao.chen.math@gmail.com}

\author{Chenyu Jin}
\address{Max Planck Institute of Dynamics and Self-Organisation, Am Fa\ss berg 17, 37077 G\"ottingen, Germany}
\email{chenyu.jin@ds.mpg.de}

\begin{document}


\begin{abstract}
	It is commonly considered that the frustration between the curvature energy
	and the chain stretching energy plays an important role in the formation of
	lyotropic liquid crystals in bicontinuous cubic phases.  Theoretic and
	numeric calculations were performed for two extreme cases: Parallel surfaces
	eliminate the variance of the chain length; constant mean curvature surfaces
	eliminate the variance of the mean curvature.  We have implemented a model
	with Brakke's Surface Evolver which allows a competition between the two
	variances.  The result shows a compromise of the two limiting geometries.
	With data from real systems, we are able to recover the G--D--P phase
	sequence which was observed in experiments.

	\textbf{Keywords:} Amphiphilic systems, Lyotropic liquid crystals,
	Bicontinuous cubic phases, Triply periodic minimal surfaces, Surface Evolver,
	Geometrical frustration 
\end{abstract}

\maketitle

\section{Introduction}\label{sec:Intro}

Liquid crystals are called \emph{lyotropic} if they experience phase
transitions by adding or removing a solvent~\cite{hiltrop1994lyotropic}.
Typically, the solute in a lyotropic liquid crystal (LLC) is amphiphilic,
comprising a hydrophilic head-group and a hydrophobic chain (tail).  Upon
varying solvent content and/or temperature, LLC can display a rich variety of
phases, including the lamellar phase, the hexagonal phase, and the bicontinuous
or micellar cubic phases.  Of these, the bicontinuous cubic phases are the most
complex and interesting.  They are observed in cells and organelles, and are
believed to play important role in biological processes
\cite{almsherqi2006cubic}.  It is generally believed that the bicontinous LLC
cubic phases can be described by triply-periodic minimal surfaces (TPMS's)
\cite{scriven1976equilibrium, larsson1980structural, longley1983bicontinuous,
mackay1985periodic}.

A LLC can be treated as a packing of curved amphiphile monolayers and water
layers.  According to the curvature of the monolayers (measured at the neutral
interfaces \cite{kozlov1991elastic, templer1995area}), LLC can be classified
into two types \cite{luzzati1968structure, seddon1995polymorphism,
hyde1989microstructure}: With respect to normal vectors pointing from oil to
water, the monolayers have positive mean curvature in a \emph{normal}
(a.k.a.~type I, oil-in-water) system, and negative mean curvature in an
\emph{inverse} (a.k.a.~type II, water-in-oil) system.  In the case of a typical
amphiphile-water binary LLC system, an inverse bicontinuous cubic phase
consists of a pair of monolayers tail-to-tail (a bilayer) draped over the TPMS
\cite{shearman2007calculations, shearman2010towards, schroder2006bicontinuous},
separating two water channels; while in a normal phase, a water layer following
the TPMS separates two channels packed by the hydrophobic tails of the
amphiphiles \cite{schroder2006bicontinuous}.  

\medskip

It is commonly believed that LLC phases arise from the competition between two
geometric demands~\cite{helfrich1973elastic,sadoc1986frustration,
anderson1988geometrical}: uniform curvature and uniform chain length in the
monolayers.  Indeed, as the amphiphilic molecules in the monolayers are
chemically identical, it is conceivable that they have the same spontaneous
curvature~\cite{hyde1990curvature} and relaxed tail length.  However, for most
of the LLC phases, the curvature and the chain length can not be simultaneously
uniform~\cite{duesing1997quantifying, anderson1988geometrical}.  Hence the
origin of \emph{frustration}, which can be quantified by the variances of mean
curvature and of chain length.

Considering both variances at the same time is difficult, hence most
investigations treat them separately.  The neutral interfaces are either
modeled as parallel surfaces of the TPMS (e.g.~\cite{anderson1988geometrical,
		templer1998gaussian,  hyde1989microstructure, templer1998modeling,
schwarz2000stability, schwarz2001bending}), eliminating the variance of the
chain length; then the Helfrich energy is calculated or measured as the energy
of the system.  Or, alternatively, the neutral interfaces are assumed to have
constant mean curvature (CMC); then the Hooke energy is used as the energy of
the system (e.g.~\cite{anderson1988geometrical, shearman2007calculations,
shearman2010towards, templer1998modeling, schwarz2000bi}).  The parallel
surface model is certainly more straight forward to calculate, but the CMC
model makes more sense for the inverse hexagonal and micellar
phases~\cite{anderson1988geometrical, duesing1997quantifying}, and is arguably
more successful in explaining the phase behaviour involving bicontinuous cubic
phases~\cite{shearman2010towards}.

There has been a constant and strong demand~\cite{anderson1988geometrical,
shearman2007calculations} for a model that allows the competition between the
curvature energy and the chain stretching energy.  Here we present a way of
modeling LLC structures in Brakke's Surface Evolver~\cite{brakke1992surface}
that fulfills this demand.  We first demonstrate our method on the inverse
hexagonal phase, then apply it to the bicontinuous cubic phases, both inverse
and normal, with the geometry of G (gyroid, $Ia3d$), D (diamond, $Pn3m$), P
(primitive, $Im3m$) TPMS's.  This gives, for the first time, a geometry of LLC
that compromises the two extremes: the parallel surfaces and the CMC surfaces.

\subsection*{Acknowledgement}

The authors appreciate discussions with Karsten Gro\ss e-Brauckmann and Ken
Brakke, and thank Gerd Schr\"oder-Turk and Lu Han for feedbacks.  Jin
acknowledges the support of Corinna C.\ Maa\ss\ and Stephan Herminghaus.

\section{Model}

Following~\cite{shearman2010towards}, the surface averaged free energy per
hydrophobic chain, denoted by $\mu$, comprises two parts:
\begin{equation}
	\mu = \mu_C + \mu_L,
\end{equation}
where
\[
	\mu_C = A (2 \kappa_H \langle (H-H_0)^2 \rangle + \kappa_G \langle K \rangle)
\]
is the Hilfrich energy or curvature energy~\cite{helfrich1973elastic,
fogden1991bending}, and
\[
	\mu_L = \kappa_L \langle (L-L_0)^2 \rangle
\]
is the Hooke energy or chain stretching energy.  Here, $A$ denotes the
cross-sectional area of a hydrophobic chain, $L$ denotes the chain length, $H$
is the interfacial mean curvature, $K$ is the interfacial Gaussian curvature,
$H_0$ is the spontaneous mean curvature, $L_0$ denotes the relaxed chain
length, and $\kappa_H$, $\kappa_G$ and $\kappa_L$ denote the moduli for the
energetic contributions from, respectively, the mean curvature, Gaussian
curvature, and hydrophobic chain stretching.  All these quantities should be
measured on or from the \emph{neutral interface}, which is the location within the
monolayer where the area is invariant upon isothermal bending
\cite{kozlov1991elastic, templer1995area}.  The average is over the whole
surface, that is $\langle x \rangle = \int_S x dS / \int_S dS$.  

The contribution from the mean curvature can be rewritten as
\begin{equation} \label{eq:sigmah}
	2A\kappa_H \left< (H-H_0)^2 \right> = 2A\kappa_H \left[ (\langle H^2 \rangle - \langle H \rangle^2) + (\langle H \rangle - H_0)^2 \right] = 2A\kappa_H \sigma_H^2 + \bar \mu_H
\end{equation}
where $\sigma_x = \langle x^2 \rangle - \langle x \rangle^2$ denotes the squared
variance.  Similarly, the contribution from the chain stretch can be decomposed
into
\begin{equation} \label{eq:sigmal}
	\mu_L = \kappa_L \left[ (\langle L^2 \rangle - \langle L \rangle^2) + (\langle L \rangle - L_0)^2 \right] = \kappa_L \sigma_L^2 + \bar \mu_L.
\end{equation}
The frustration of the system is measured as a weighted sum of the squared variances
\[
	2A\kappa_H \sigma_H^2 + \kappa_L \sigma_L^2,
\]
which is the quantity to minimize in our model.

This model is certainly a simplification.  In particular, we ignore the
contribution of the tilt energy (which will be justified later), the higher
order contributions of the curvatures, as well as the non-local interactions of
the monolayers, such as van der Waals interaction and hydration repulsion.
Nevertheless, it is hoped that the important physical features are retained.

\medskip

Brakke's Surface Evolver~\cite{brakke1992surface} is a software that minimizes
energies of triangulated surfaces subject to constraints and boundary
conditions.  Apart from the usual surface tension energy, or the area
functional, Surface Evolver is able to calculate many other quantities on a
surface.  Many of these quantities can be included in the energy to minimize.
A quantity can be calculated on a particular set of geometric elements
(vertices, edges, faces, bodies), allowing the users to control which elements
contribute to the total energy.  Moreover, each quantity has a modulus to
specify its weight in the total energy.  The moduli are adjustable in real time
on each geometric element.

In particular, Surface Evolver has implemented the Willmore energy $(h-h_0)^2$
(see~\cite{hsu1992}) evaluated on vertices, and the Hooke energy
$(\ell-\ell_0)^2$ evaluated on edges.  Here, we use lower case letters to
distinguish dimensionless quantities in Surface Evolver from physical
quantities.  More specifically, our computations in Surface Evolver assume unit
lattice parameter (edge length of the conventional cubic cell of the TPMS).  If
the physical system has lattice parameter $a$, we have the relations $h=Ha$,
$k=Ka^2$ and $l=L/a$.  Then the surface averaged energy per hydrophobic chain
takes the following form
\begin{equation}
	\mu = Aa^{-2} \left[ 2 \kappa_H \langle (h-h_0)^2 \rangle + \kappa_G \langle k \rangle \right] + \kappa_L a^2 \langle (\ell-\ell_0)^2 \rangle.
\end{equation}

The parameters $h_0$ and $\ell_0$ in Surface Evolver are supposed to mean the
dimensionless spontaneous mean curvature and relaxed chain length.  However, we
keep updating these parameters to the average values $\langle h \rangle$ and
$\langle \ell \rangle$, so that the averaged Willmore energy gives the squared
variance of mean curvature, and the averaged Hooke energy gives the squared
variance of chain length.  Hence the minimized quantity is the frustration
\[
	2A\kappa_Ha^{-2}\sigma_h^2 + \kappa_L a^2 \sigma_\ell^2
\]
or, for convenience,
\[
	\sigma_h^2 + \lambda\sigma_\ell^2
\]
where $\lambda = \kappa_L a^4 / 2A\kappa_H$.  After sufficiently evolving the
surface, it is easy to transform the minimized frustration into the energy
using Eqs.~\eqref{eq:sigmah} and~\eqref{eq:sigmal}.  

Our procedure of modeling bicontinuous LLC cubic phases in Surface Evolver can
be outlined in the following five steps.

\begin{enumerate}
	\item Prepare a triangulation of the D, P or G surface, or any other surface
		of interest.

	\item Make two copies of the triangulation, modeling the neutral
		interfaces.  Assign Willmore energy to these copies.

	\item Create edges modeling the hydrocarbon chains, and assign Hooke energy
		to these edges.

	\item Impose a volume constraint to the space between the neutral interfaces
		in correspondance with the desired water fraction.

	\item Evolve the surface towards the minimum of the total energy, while keep
		updating the average values $h_0 = \langle h \rangle$ and $\ell_0 = \langle
		\ell \rangle$.
\end{enumerate}

In step (3), the edges modeling the hydrophobic chains depend on the system and
the model.  For inverse LLC phases, we are aware of several ways of modeling the chains:

\renewcommand{\theenumi}{\alph{enumi}}
\begin{enumerate}
 	\item Connect edges between the corresponding vertices on the triangulated
 		neutral interfaces.

 	\item Connect edges from the vertices on the neutral interfaces to the
 		corresponding vertices on the TPMS.

 	\item Subtending normal vectors from the TPMS until the neutral interfaces
 		(used in~\cite{shearman2007calculations}).

 	\item Subtending normal vectors from the neutral interfaces until the
 		TPMS.
\end{enumerate}

In order to consider the curvature and the chain length at the same time, we
need to keep the chains up to date at every step of surface evolution.  The
models (c) and (d) require computing the intersection point of a line and a
triangulated surface, which is very time consuming.  We will use the models (a)
and (b).  The chains are established once at the beginning, and will be used
during the entire computation.

However, only the model (d) forces the chains to be perpendicular to the
neutral interfaces.  This may raise concerns on the legitimacy of ignoring tilt
energy (see \cite{hamm2000elastic}) in other models.  We address to this
concern by measuring the amount of tilting.  It turns out that the chains do
not tilt much, at least in the inverse bicontinuous cubic phases, thanks to the
\texttt{normal\_motion} mode of the Surface Evolver.

An edge in the model (a) corresponds to two chains.  One advantage of this
model is that it is independent of the TPMS: the TPMS is present only for
reference, and does not participate in the calculation.  Our program is robust
in the sense that, if one starts from a periodic CMC surface instead of a TPMS
(which the author did, accidentally), the bilayer will correctly evolve to the
TPMS.  Hence our program confirms, under the current theory, that bicontinuous
LLC cubic phases do follow the geometry of TPMS.

\begin{figure}[p]
  \centering
  \includegraphics[width=\textwidth]{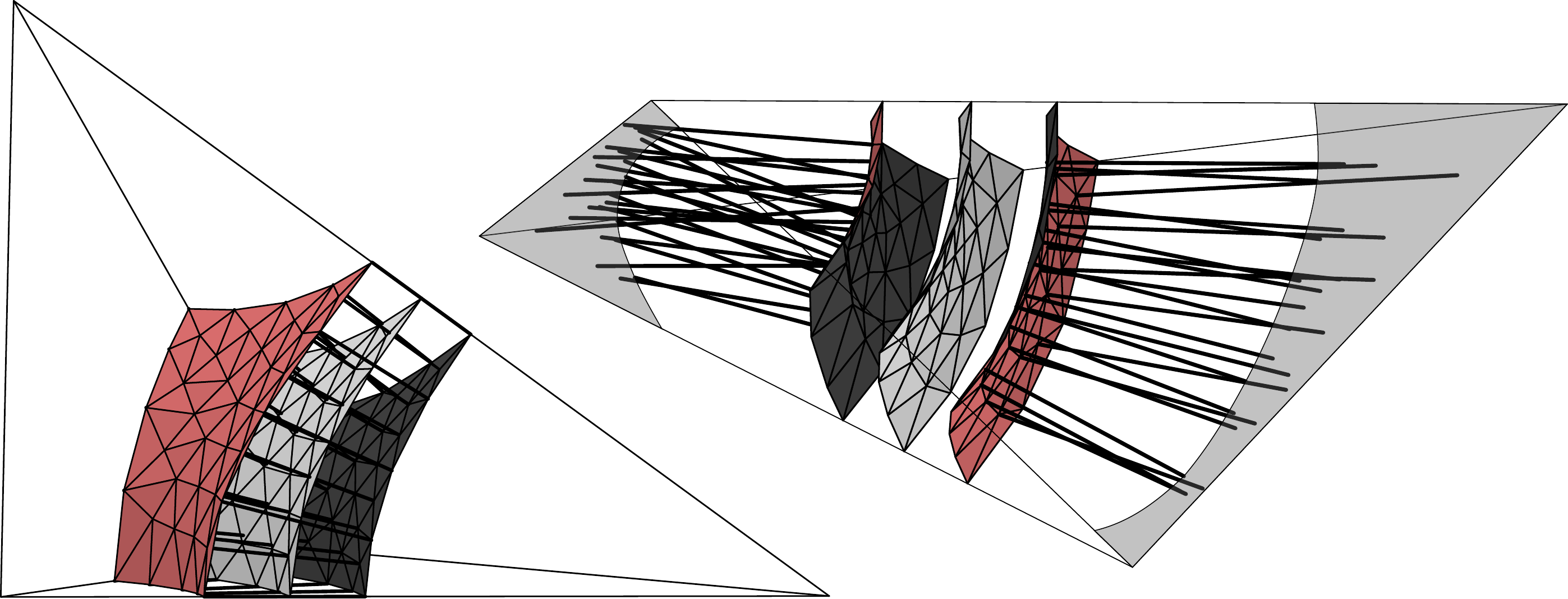}
  \caption{
  	Our Surface Evolver models illustrated for the P phase.  A triangulation of
  	the P patch (light grey) is copied twice.  The copies (red and dark grey)
  	model the neutral interfaces.  The black bold segments are the edges
  	modeling the hydrophobic chains.  Left: model (a) for the inverse LLC P
  	phase, with hydrophobic chains connecting corresponding vertices on the
  	neutral interfaces.  Right: model for the normal LLC P phase, with
  	hydrophobic chains connecting the neutral interfaces to the medial
  	surface (grey area).  \label{fig:model}
	}
\end{figure}

For normal LLC phases, we subtend normal vectors from vertices of the
triangulated TPMS to find intersection points on the medial surface, then model
the hydrophobic chains by edges from these intersection points to the
corresponding vertices on the neutral interfaces.  This model follows the idea
of~\cite{schroder2003medial}.  See Figure~\ref{fig:model} for an illustration
of our models.

The readers are free to use other chain models in their own calculations.  The
technical details are postponed to Section~\ref{sec:tech}.

\section{Result}

\subsection{Inverse hexagonal phase}

In the inverse hexagonal LLC phase, water is contained in the tubes formed by
amphiphiles arranged in a hexagonal lattice.  Due to its simplicity, the
inverse hexagonal phase has been a standard example
(e.g.~\cite{duesing1997quantifying, anderson1988geometrical}) to illustrate the
incompatibility between uniform curvature and uniform chain length, and to
demonstrate the calculation of chain frustration within the CMC model.

We also choose the hexagonal phase to explain our method, since the model can
be built up in dimension $2$.  More specifically, the hexagonal lattice graph
plays the role of TPMS, and a ``triangulation'' is nothing but a subdivision of
the edges.  The neutral interfaces are modeled by cycles within the hexagonal
cells.  In practice, such a cycle is initially created as a copy of the
hexagon.  Vertices of the cycles are connected to the corresponding points on
the hexagonal lattice graph, modelling the hydrophobic chains of the
amphiphiles (model (b)).  The area between the cycle and the hexagon
corresponds to half of the volume between the neutral interfaces.

Surface Evolver is able to apply squared curvature energy on the cycle, and
Hooke energy on the hydrophobic chains.  We can choose to minimize either
energy alone, or minimize a weighted sum of the two.  At each iteration, we
update the parameter $\ell_0$ to the average chain length, so that the averaged
Hooke energy actually gives the dimensionless squared variance of chain length.
Note that the average curvature is a constant here.

\begin{figure}[p]
  \centering
  \includegraphics[width=\textwidth]{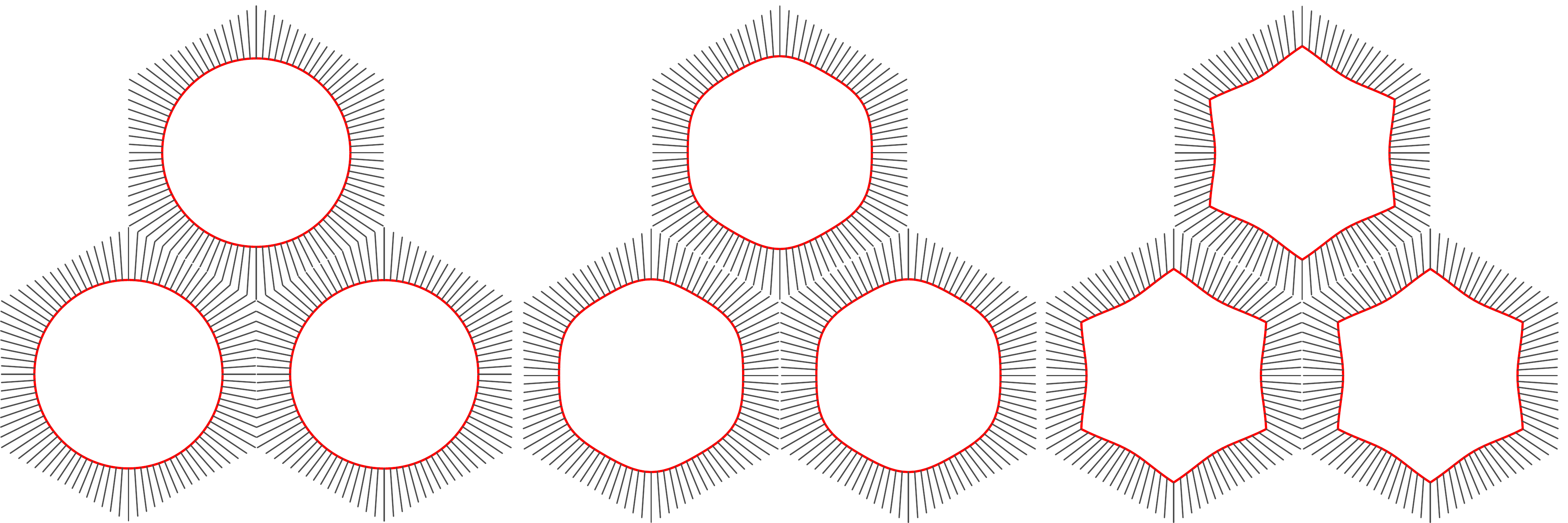}
  \caption{
  	Evolved results of our model applied to the inverse hexagonal phase by
  	minimizing only the curvature energy (left), only the chain stretching
  	energy (right), and the frustration as a weighted sum of the two (middle).
  	The cycles in red model the neutral interfaces, and the segments in grey
  	model the hydrophobic chains (model (b)).\label{fig:hex}
  }
\end{figure}

In Figure~\ref{fig:hex}, we show the result of Surface Evolver under three
different situations.  Minimizing curvature energy alone results in a
circle\footnote{This is related to a quite challenging mathematical problem;
see~\cite{ferone2015}.}, and minimizing chain stretching energy alone results in a
star-like shape.  If both are present in the total energy, a competition
arise.  The cycle evolves to a closed curve, which is neither a circle nor a
star, but an intermediate shape.

\subsection{Inverse LLC phases}

We now repeat the calculation of Shearman et
al.~\cite{shearman2007calculations} to verify the feasibility of our method.
More specifically, we measure the chain frustration in inverse LLC phases
assuming CMC neutral interfaces.  The hydrophobic chains are modeled by edges
connecting corresponding vertices on the triangulated neutral interfaces;
recall that they are copies of the same triangulation.  The chain lengths are
then half of the edge lengths.  The volume fraction $\phi_n$ of the space
between the neutral interfaces ranges from $0.01$ to $0.50$.  The squared
variance $\sigma_\ell^2$ of dimensionless chain length is measured as the
frustration of the system.

\begin{figure}[p]
	\includegraphics[width=.8\textwidth]{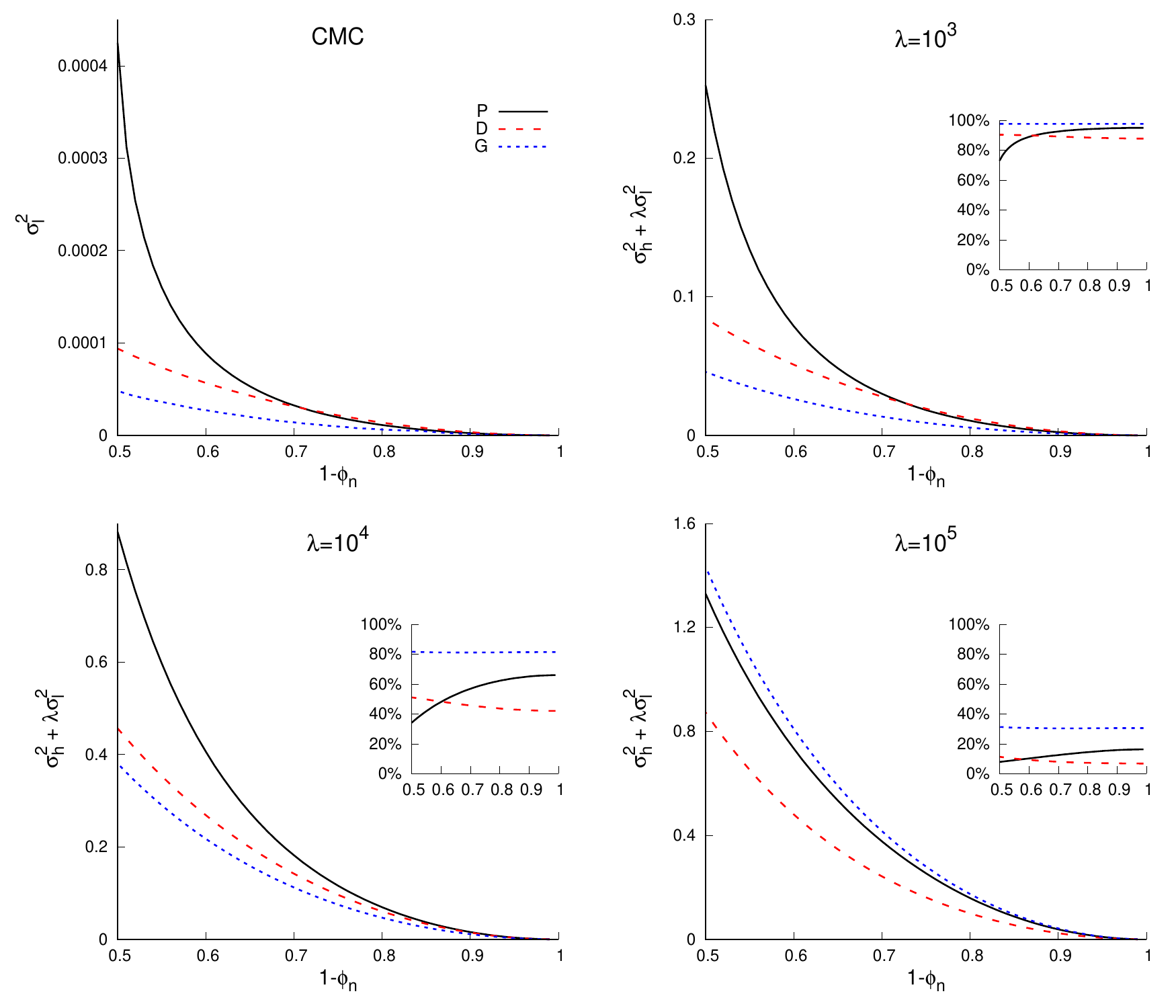}
	\caption{
		Frustration of the inverse D, P and G phases against $1-\phi_n$.  Top-left:
		squared variance of chain length of the CMC model, to compare
		with~\cite{shearman2007calculations}.  Other: Weighted sum of the squared
		variances with $\lambda=10^3$, $10^4$ and $10^5$.  The insets show the
		contribution of the squared variance of chain length in the total
		frustration. \label{fig:inverse}
	}
\end{figure}

The result is plotted in the top-left panel of Figure~\ref{fig:inverse}.  We
use $1-\phi_n$ on the $x$-axis to facilitate the comparison with Figure~5
in~\cite{shearman2007calculations}.\footnote{The ``$\phi_w$''
	in~\cite{shearman2007calculations} should be $1-\phi_n$.  The relation
between $\phi_n$ and $\phi_w$ is $\phi_n=(1-\phi_w)\nu_n/\nu$, where $\nu$ is
the molecular volume and $\nu_n$ is the volume between the neutral interface to
the chain ends.} The two plots are very similar:  The frustration is the lowest
in the G phase, and the highest in the P phase when the water fraction is low,
or in the D phase when the water fraction is high.  In particular, we also
observe a crossover between the P and D phases around $\phi_n=0.29$ ($\approx
0.25$ in ~\cite{shearman2007calculations}).  There is only a slight numerical
disagreement, which can be explained by the minor differences in our models.
In particular, the hydrophobic chains in~\cite{shearman2007calculations} is
perpendicular to the TPMS\footnote{To be rigorous, the chains should be
perpendicular to the CMCs}, while we allow slight tilting for the chains.

Now that we have confirmed the validity of our algorithm, we turn on the Hooke
energy in Surface Evolver, and minimize the weighted sum of the two
dimensionless squared variances.  We fix the weight of $\sigma_h^2$ to be $1$,
and vary the weight of $\sigma_l^2$.  In other words, we are minimizing
$\sigma_h^2 + \lambda \sigma_\ell^2$, corresponding to the frustration divided
by $2A\kappa_Ha^{-2}$.  In view of parameter values provided
in~\cite{shearman2010towards}, we choose $\lambda=10^3$, $10^4$, $10^5$ in our
computations; see context of Figure~\ref{fig:exp}. The result surface is
neither parallel nor constant mean curvature, but a compromise of the two.  The
measured frustrations, together with the contribution of chain length
frustration, is plotted in Figure~\ref{fig:inverse}.  Histograms in
Figure~\ref{fig:tilt-inv} show tilt angles less than $10$ degrees, justifying
our practice of ignoring the tilt energy.  

The mean curvature and the chain length compete to be uniform, and a higher
weight is an advantage in this competition, hence the value of $\sigma_\ell^2$
is lower comparing to the CMC model.  Moreover, the contribution of
$\lambda\sigma_\ell^2$ in the total frustration is decreasing as $\lambda$
increases.  This trend is more significant in the D and P phases than in the G
phase.  As a consequence, the frustration in the G phase eventually surpasses
the D and P phases when $\lambda = 10^5$.

\begin{figure}[p]
  \centering \includegraphics[width=.8\textwidth]{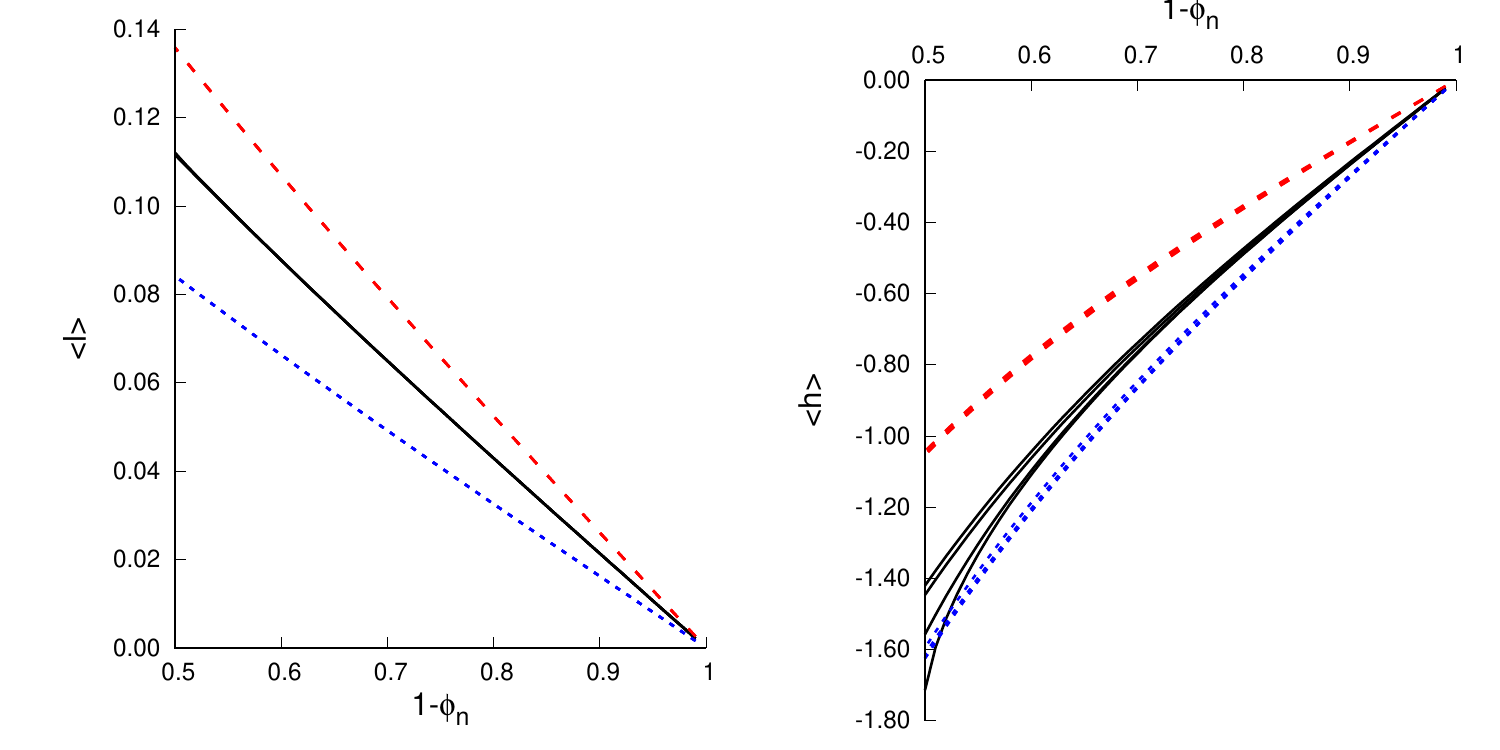}
  \caption{
  	The dimensionless average chain length $\left< \ell \right>$ (left) and
  	average mean curvature $\left< h\right>$ (right) of inverse bicontinuous
  	cubic phases. The plots with CMC, $\lambda=10^3$, $\lambda=10^4$ and
  	$\lambda=10^5$ are overlapped to emphasize on their similarity.
  } \label{fig:meanHL-inv}
\end{figure}

\begin{figure}[p]
  \centering \includegraphics[width=.7\textwidth]{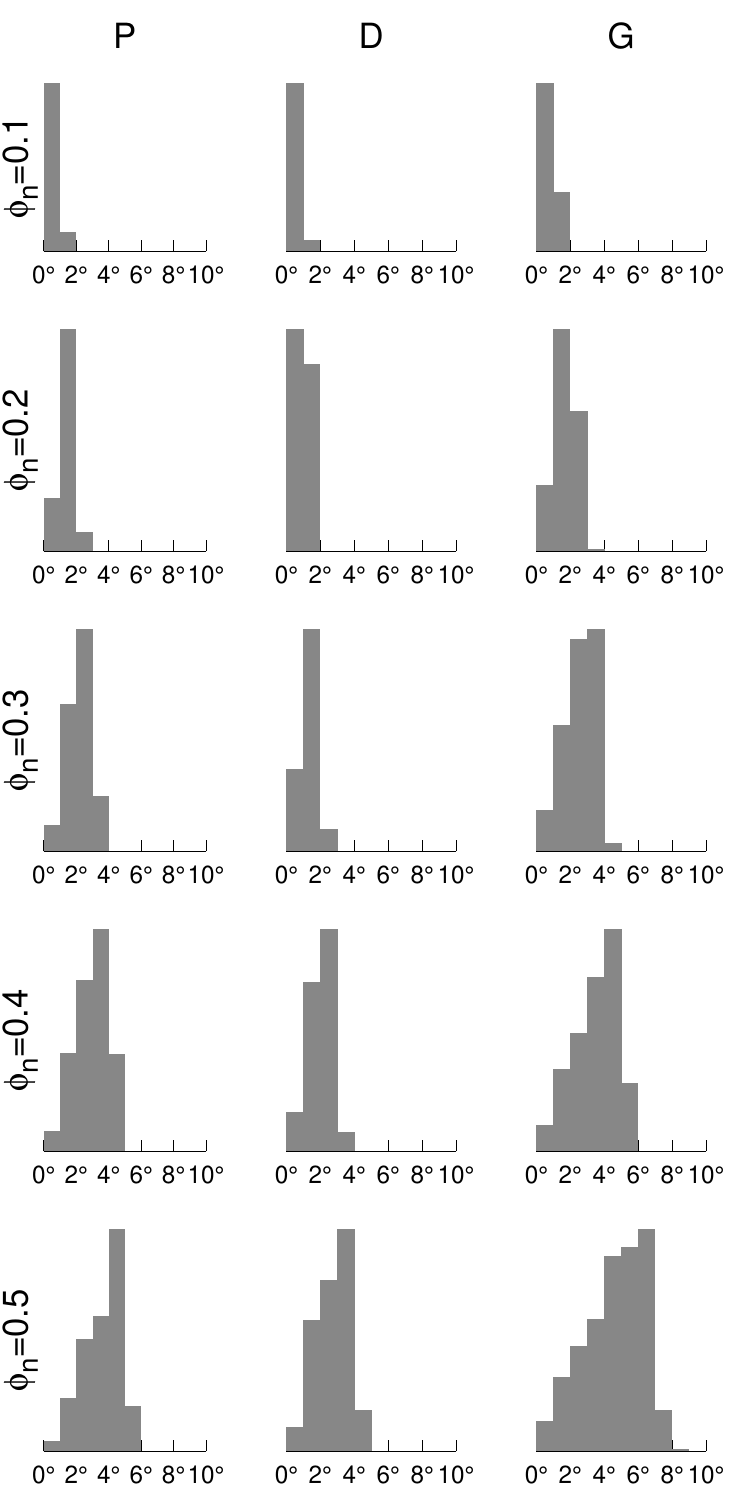}
  \caption{Histograms of chain tilt angles in the inverse bicontinuous cubic
  phases with different $\phi_n$.} \label{fig:tilt-inv}
\end{figure}

\begin{figure}[p]
  \centering
  \includegraphics[width=.8\textwidth]{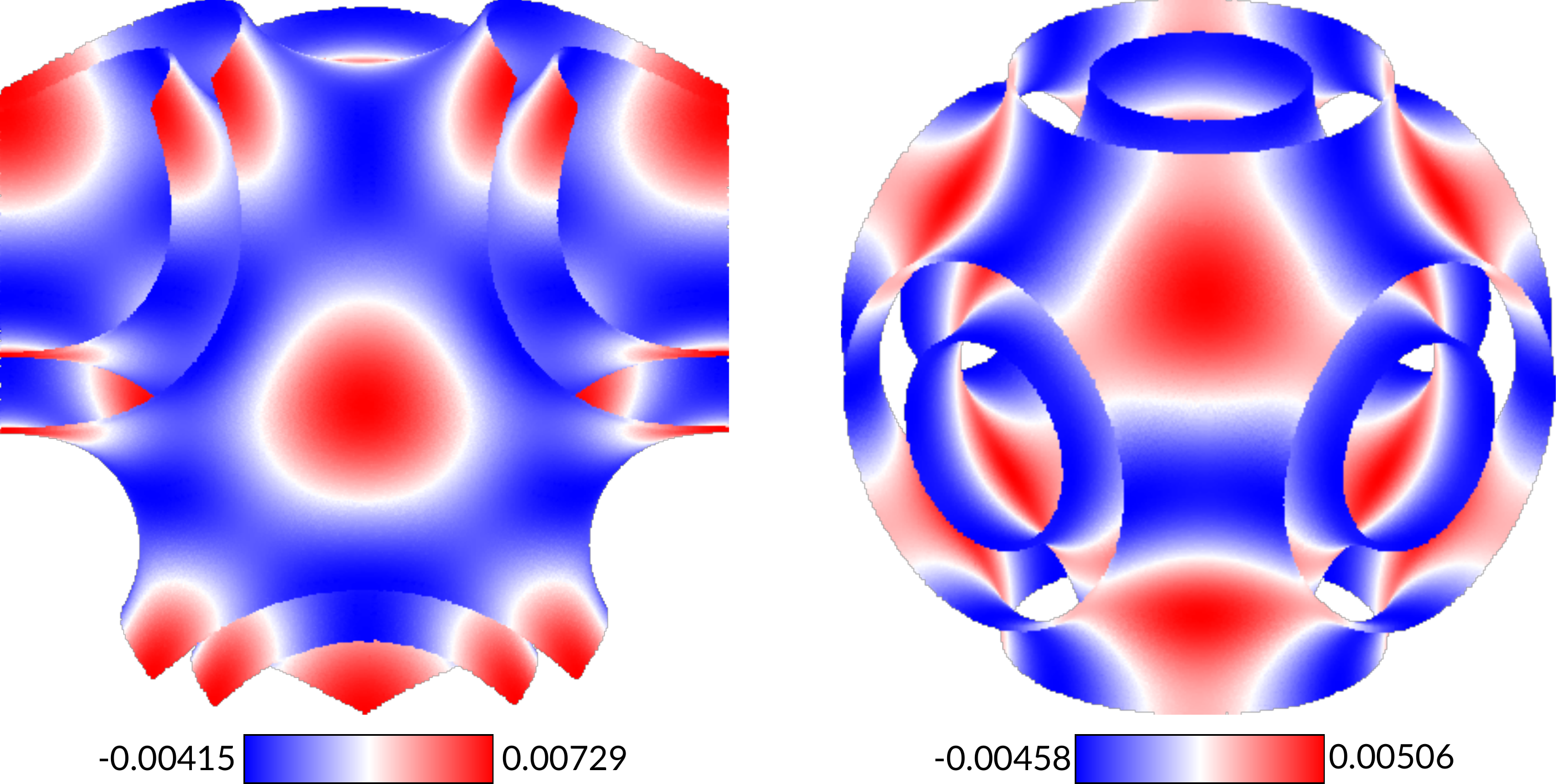}

  \includegraphics[width=.8\textwidth]{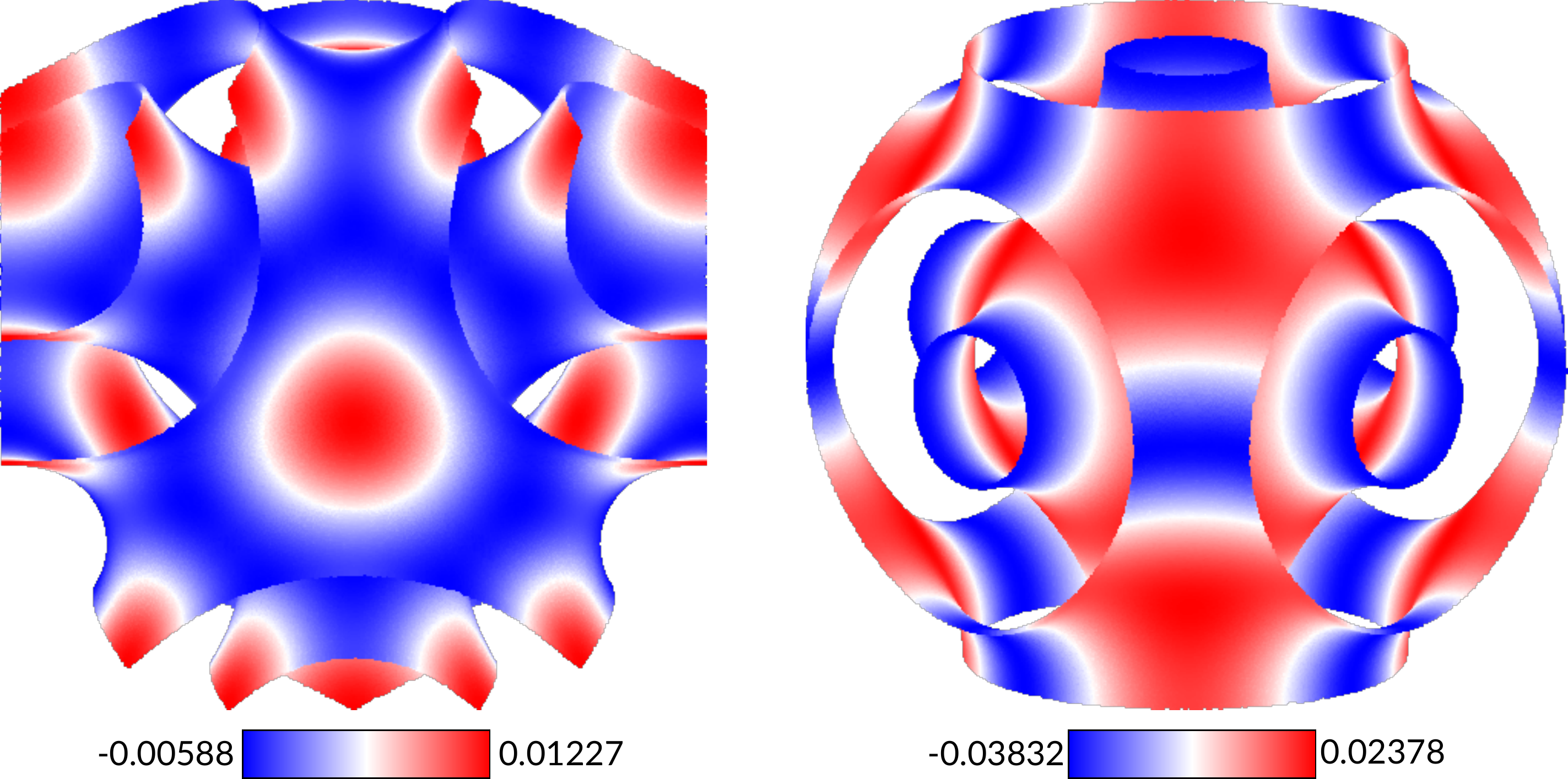}
  \caption{
  	Evolved neutral interfaces in the inverse D (left) and P (right) phases
  	with $\lambda=10^4$, and the space between the neutral surfaces, which is
  	completely occupied by hydrophobic chains, has a volume fraction
  	$\phi_n=0.3$ (top) or $\phi_n=0.5$ (bottom).  The color shows the
  	difference from the CMC surface with the same $\phi_n$: red indicates a
  	deviation away from the TPMS, and blue indicates a deviation towards the
  	TPMS.  The color bar below each figure shows maximum deviations, normalized
  	with unit lattice parameter.  Clearly, the competition causes expansions in
  	the necks.\label{fig:DiffInv}
	}
\end{figure}

\begin{figure}[p]
  \centering \includegraphics[width=.8\textwidth]{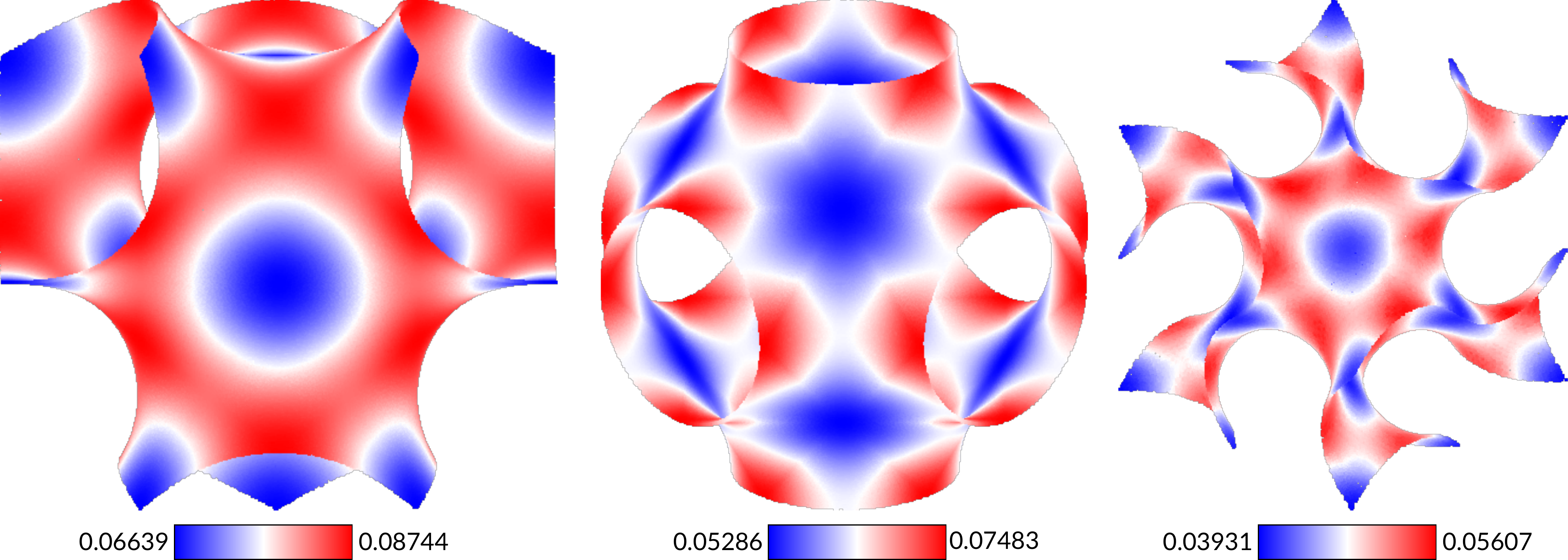}
  \caption{
  	The D(left), P(middle) and G(right) TPMS color by its distance to the CMC
  	neutral surfaces with $\phi_n=0.3$.  The color bar shows the minimum and
  	maximum distance, normalized with unit lattice parameter.  This plot is a
  	reproduction of Figure~8 of ~\cite{shearman2007calculations}.
	} \label{fig:cmc}
\end{figure}

\begin{figure}[p]
	\includegraphics[width=.8\textwidth]{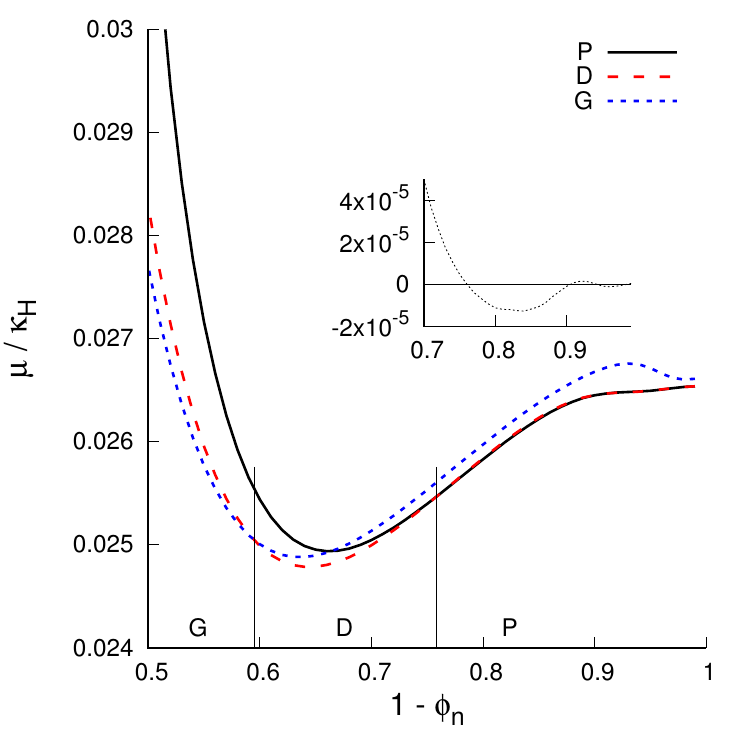}
	\caption{
		Surface averaged energies per hydrophobic chain divided by $\kappa_H$ of
		inverse D, P and G are plotted against $1-\phi_n$, with $A=33
		\text{\AA}^2$, $\kappa_G/\kappa_H = -0.75$, $\kappa_L/\kappa_H = 0.00035
		\text{\AA}^{-2}$, $H_0=1/62.8 \text{\AA}^{-1}$ and $L_0=8.8\text{\AA}$.
		With the increasing $1- \phi_n$, the phase with the lowest energy gives the
		sequence G--D--P.  The inset shows the energy difference of P over D at
		$1-\phi_n$ larger than $0.7$.
	} \label{fig:exp}
\end{figure}

As an example, the evolved P and D neutral interfaces with $\phi_n=0.3$ or
$0.5$ and $\lambda=10^4$ are shown in Figure~\ref{fig:DiffInv}.  The difference
of the resulting neutral interfaces from the CMC interfaces with the same
volume fraction $\phi_n$ is not visible with human eyes.  This is also observed
in Figure~\ref{fig:meanHL-inv} where plots of average chain length and average
mean curvature are overlapped to show the similarity.  We then use
CloudCompare~\cite{cloudcompare} to measure the deviation from the CMC
interfaces, and color the surface accordingly for visualization.  Roughly
speaking, comparing to the CMC model, the neutral interfaces tend to move away
from the TPMS around the flat points (where Gaussian curvature vanishes) of the
TPMS, and towards the TPMS at other places.  This is compatible with the
observation in the CMC model that the chains are compressed around the flat
points of the TPMS, and extended other where; see Figure~8
of~\cite{shearman2007calculations}, or Figure~\ref{fig:cmc} for a colorful
reproduction.

If the water fraction decreases further, the neutral interfaces in the CMC
model tends to spheres (positive Gaussian curvature) connected by small necks
(negative Gaussian curvature)~\cite{anderson1987periodic, grosse2012triply}.
Eventually, there will be a pinch-off point due to certain physical threshold,
e.g., molecular size, and the bicontinuous phases are substituted by micellar
phases.  For the P phase, the pinch-off is supposed to occur at
$\phi_n=0.5$~\cite{shearman2007calculations}.  This can be observed in the
top-left panel in Figure~\ref{fig:inverse}, as the curve for the P phase
becomes very sloped near $\phi_n=0.5$.  However, water fraction in experiment
can be much smaller~\cite{shearman2010towards}.  This inconsistency is a weak
point of the CMC model.  In our model, the plot curve becomes less sloped,
implying that the pinch-off point is moving towards lower water fraction.
Indeed, we see in Figure~\ref{fig:DiffInv} that the competition causes
expansions in the necks, therefore delays the pinching off.  Hence our model is
able to cover highly dehydrated LLCs observed in experiment, which can not be
explained by the CMC model~\cite{shearman2010towards, templer1998gaussian}.

\begin{figure}[p]
  \centering
  \includegraphics[width=\textwidth]{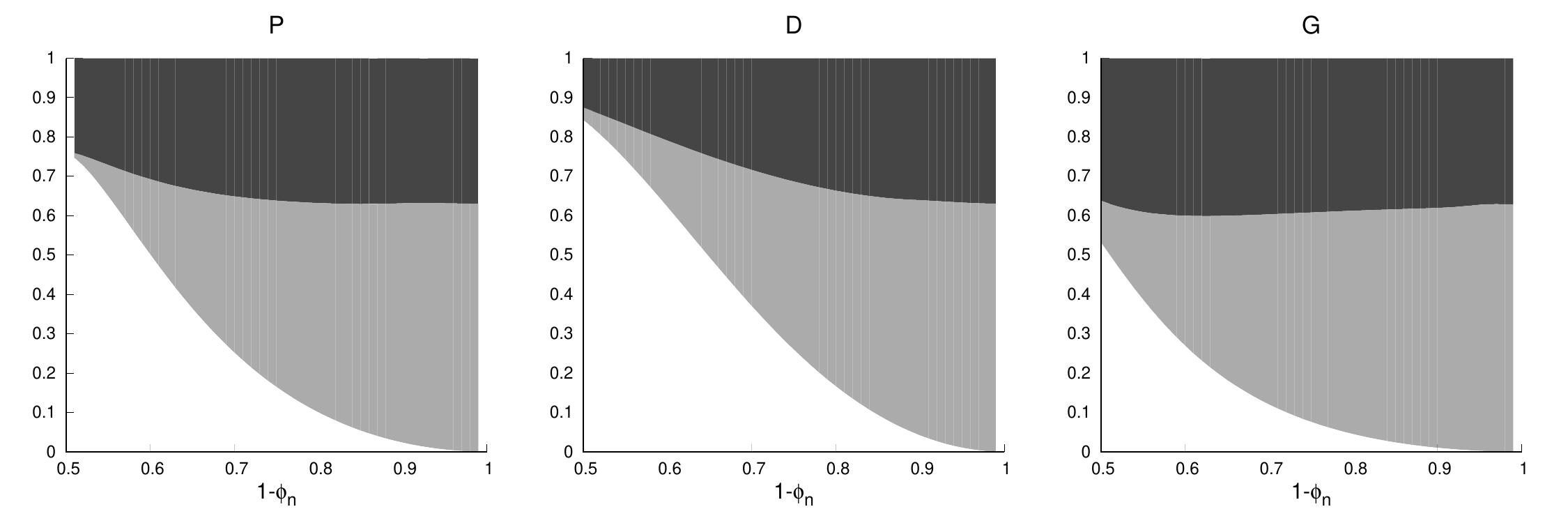}
  \caption{
  	Energy composition in the calculation based on experimental system (see
  	Figure~\ref{fig:exp}).  Different energies are indicated by colors: dark
  	grey for $\mu_L$, light grey for $2A\kappa_H \left< (H-H_0)^2 \right>$, and
  	white for $\kappa_G \langle K \rangle$.
	} \label{fig:compose}
\end{figure}

Calculations shown in Figure~\ref{fig:inverse} assume constant modulus, thus do
not correspond to any experimental systems.  To connect to the real systems, we
need to know the lattice parameter $a$.  Recall that the surface averaged
energy per hydrophobic chain divided by $\kappa_H$, expressed with
dimensionless quantities, is
\[
  \frac{\mu}{\kappa_H} = \frac{A}{a^2} \left[ 2 \left< (h-h_0)^2 \right> + \frac{\kappa_G}{\kappa_H} \left< k \right> \right] + \frac{\kappa_La^2}{\kappa_H} \left< (\ell - \ell_0)^2 \right>.
\]
The total Gaussian curvature $\int_S k dS$ is a topological constant
$2\pi\chi$, where $\chi$ is the Euler characteristic of the TPMS in the
conventional cubic unit.  The value of $\chi$ is $-4$ for the P surface, $-2$
for the D surface and $-8$ for the G surface.  A relation between $a$ and
$\phi_n$, as well as typical ranges for other parameters, are available
in~\cite{shearman2010towards}.  We take the values $A=33 \text{\AA}^2$,
$\kappa_G/\kappa_H = -0.75$, $\kappa_L/\kappa_H = 0.00035 \text{\AA}^{-2}$,
$H_0=1/62.8 \text{\AA}^{-1}$.  As for the relaxed chain length, we take
\emph{for the moment} $L_0=8.8\text{\AA}$, which is the measured distance from
the chain ends to the neutral interface in the G phase~\cite{chung1994neutral}
of $1$-monoolein/water system.  These allow us to plot $\mu /\kappa_H$ in
Figure~\ref{fig:exp} as a function of $1-\phi_n$.  The phases of lowest energy
gives a G--D--P sequence with increasing water fraction, which is a widely
observed phase sequence in experiments~\cite{tenchov1998accelerated,
tenchov2006cubic} (see also~\cite{qiu2000phase}).  The energy compositions of
different phases are plotted in Figure~\ref{fig:compose}.

However, the readers should be warned that the relation between $a$ and
$\phi_n$, as well as some parameter values we use in this calculation, is
derived under the CMC model.  Our result is indeed very close to the CMC model.
For instance, there is no visible difference in the dimensionless average chain
length; see Figure~\ref{fig:meanHL-inv}.  However, the difference may become
significant with low water fraction.  In the right panel of
Figure~\ref{fig:meanHL-inv}, the plots of average mean curvature divert near
$\phi_n=0.5$.  When $\phi_n=0.5$, the deviation of the evolved neutral
interfaces from CMC interfaces is comparable with the chain length; see bottom
of Figure~\ref{fig:DiffInv}.  Hence the usability of these parameter values
is subject to further verification.

\subsection{Normal LLC phases}

LLC in normal bicontinuous cubic phases seem to be less common.  The authors
are not aware of any reports other than normal bicontinuous G phase
\cite{sakya1997thermotropic, alexandridis1998record}.  Meanwhile, in mesophased
silica system, which can be seen as LLC systems with water replaced by silica,
various normal phases have been reported, including the bicontinuous
G~\cite{monnier1993cooperative, anderson1997simplified, anderson2005new},
P~\cite{jain2005direct} and D~\cite{gao2006synthesis} phases.

Despite of our limited understanding, we repeat the same calculations to the
normal bicontinuous LLC cubic phases, with the volume fraction $\phi_n$ of the
space between the neutral interfaces (containing water) ranging from $0.01$ to
$0.50$.  Following Schr\"oder-Turk et al.~\cite{schroder2003medial,
schroder2006bicontinuous}, the hydrophobic chains are modeled by edges from
vertices on the triangulated neutral interfaces along the normal vectors to the
nearest point on the medial surface of the TPMS.

Because of the way that Surface Evolver calculates the volume, the measurements
near $\phi_n=0$ are not physically meaningful; see Section~\ref{sec:tech}.  The
tilt angles are much larger than in the inverse phases, hence the tilt energy
is not really ignorable; see histograms in Figure~\ref{fig:tilt-nor}.  Hence we
do not attempt to connect our calculation with real system.  For these reasons,
and also due to lack of CPU power, we only compare the CMC model with the case
$\lambda=10^4$.  

\begin{figure}[p]
	\includegraphics[width=.8\textwidth]{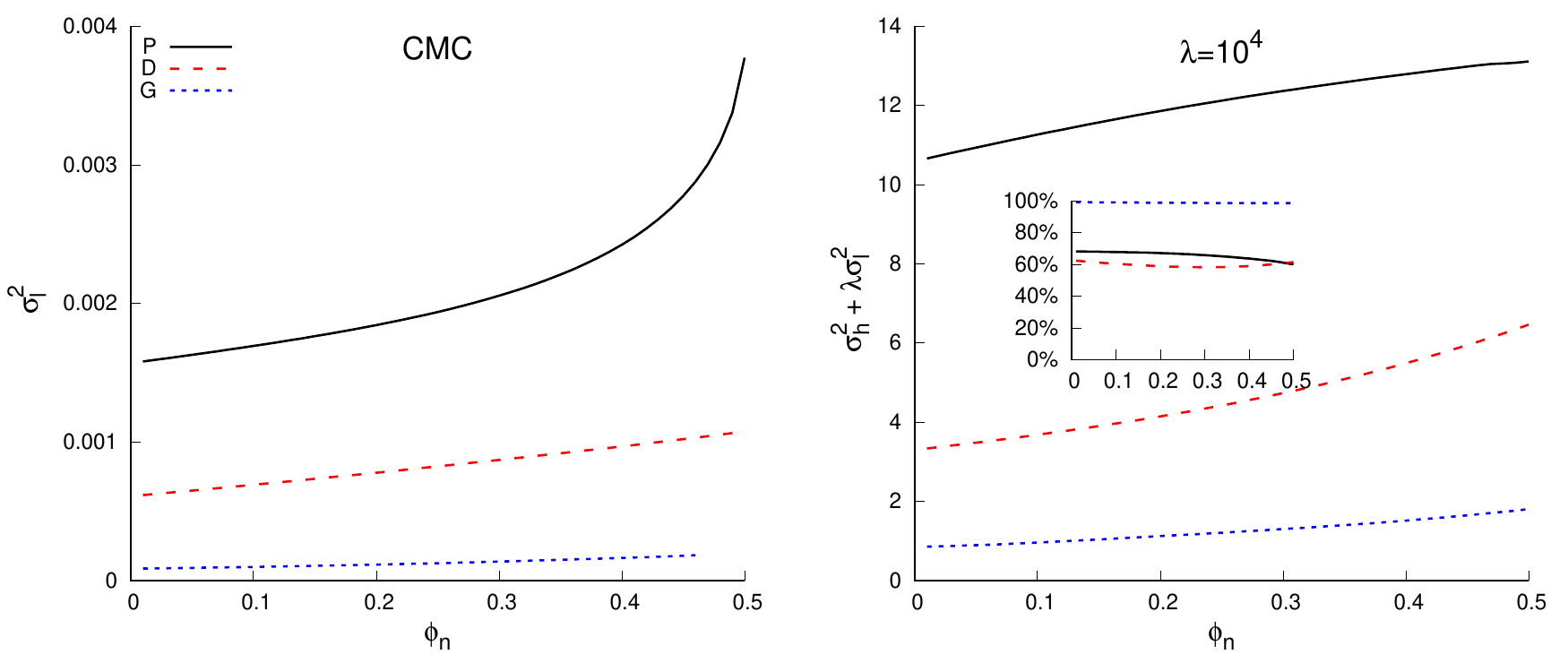}
	\caption{
		Frustration of the inverse D, P and G phases against $1-\phi_n$.  Left:
		Squared variance of chain length of the CMC model.  Right: Weighted sum of
		the squared variances of our model with $\lambda=10^4$.  The inset shows
		the contribution of the squared variance of chain length in the total
		frustration. \label{fig:normal}
	}
\end{figure}

The measured frustrations for the P, D and G phases, as well as the
contribution of chain stretching energy when $\lambda=10^4$, are plotted in
Figure~\ref{fig:normal}.  In both CMC model and our model, the frustration is
the highest in the P phase and the lowest in the G phase. This could explain
the difficulty of observing the normal bicontinuous P and D phases in
experiment. As in the inverse phases, we see that the value of $\sigma_\ell^2$,
as well as the contribution of $\lambda\sigma_\ell^2$, declines as $\lambda$
increases.  The effect is again more significant in the D and P phases than in
the G phases.

\begin{figure}[p]
  \centering \includegraphics[width=.8\textwidth]{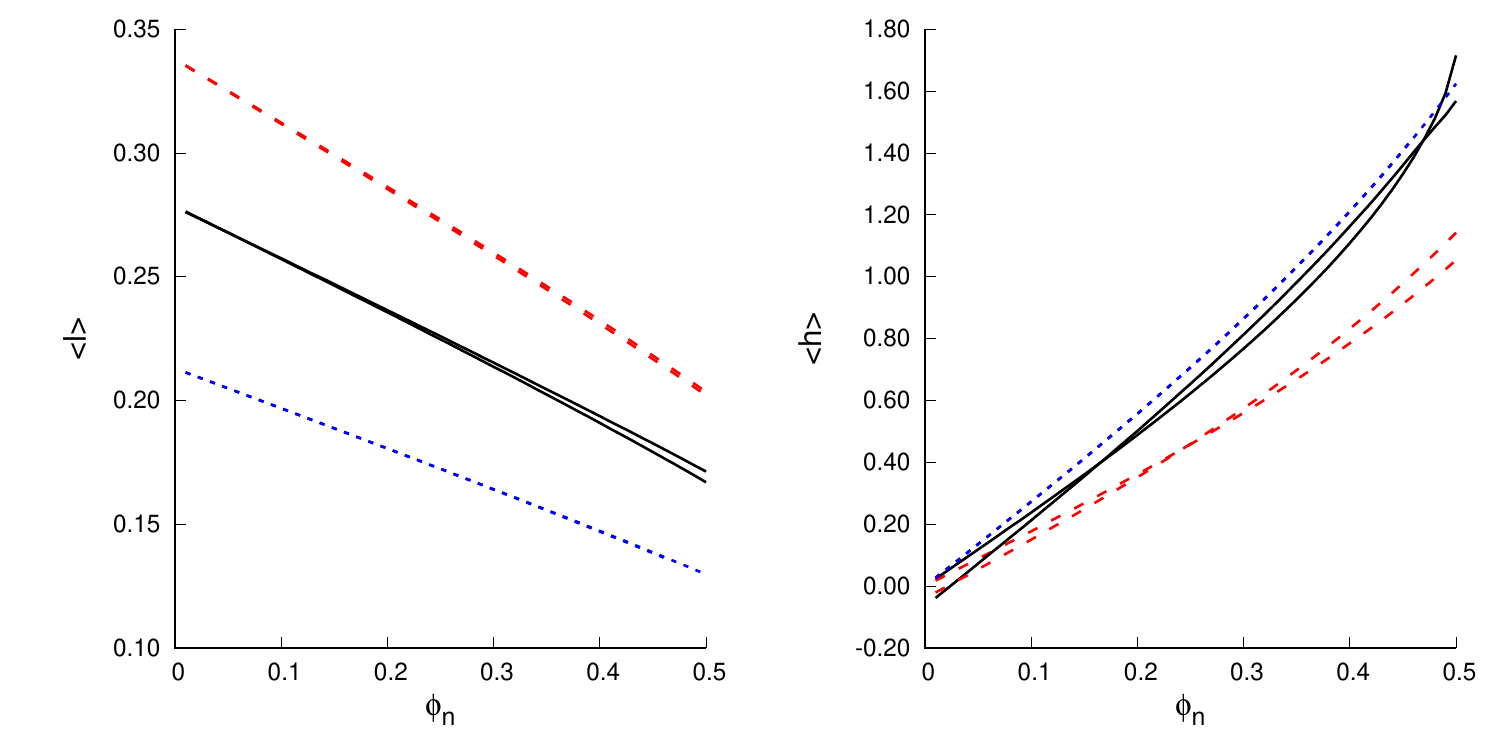}
  \caption{
  	The dimensionless average chain length $\left< \ell \right>$ (left) and
  	average mean curvature $\left< h \right>$ (right) of normal bicontinuous
  	cubic phases. The plots with CMC and $\lambda=10^4$ are overlapped to
  	emphasize on their similarity.
} \label{fig:meanHL-nor}
\end{figure}

\begin{figure}[p]
  \centering
  \includegraphics[width=.7\textwidth]{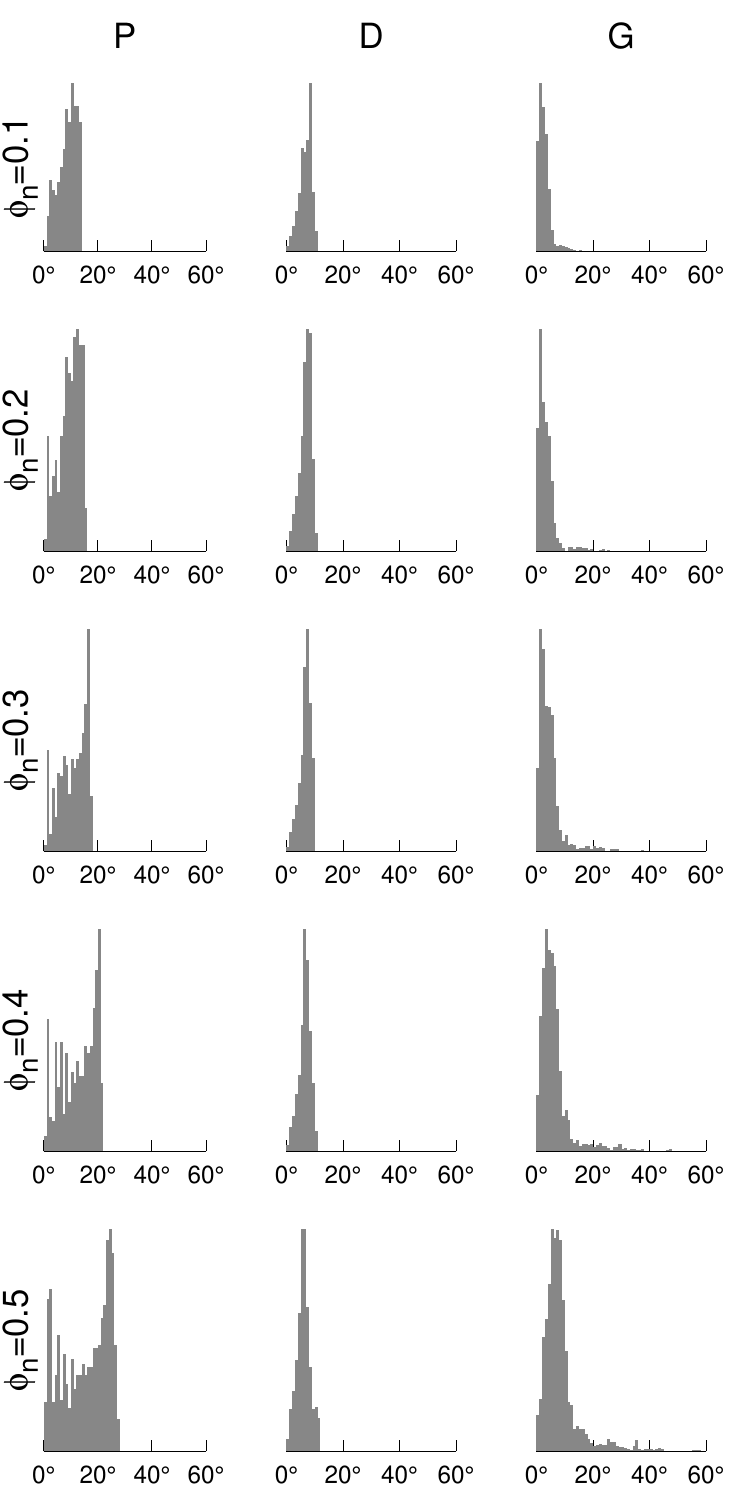}
  \caption{Histograms of chain tilt angles in the normal phases with different $\phi_n$.}
  \label{fig:tilt-nor}
\end{figure}

\begin{figure}[p]
  \centering
  \includegraphics[width=.8\textwidth]{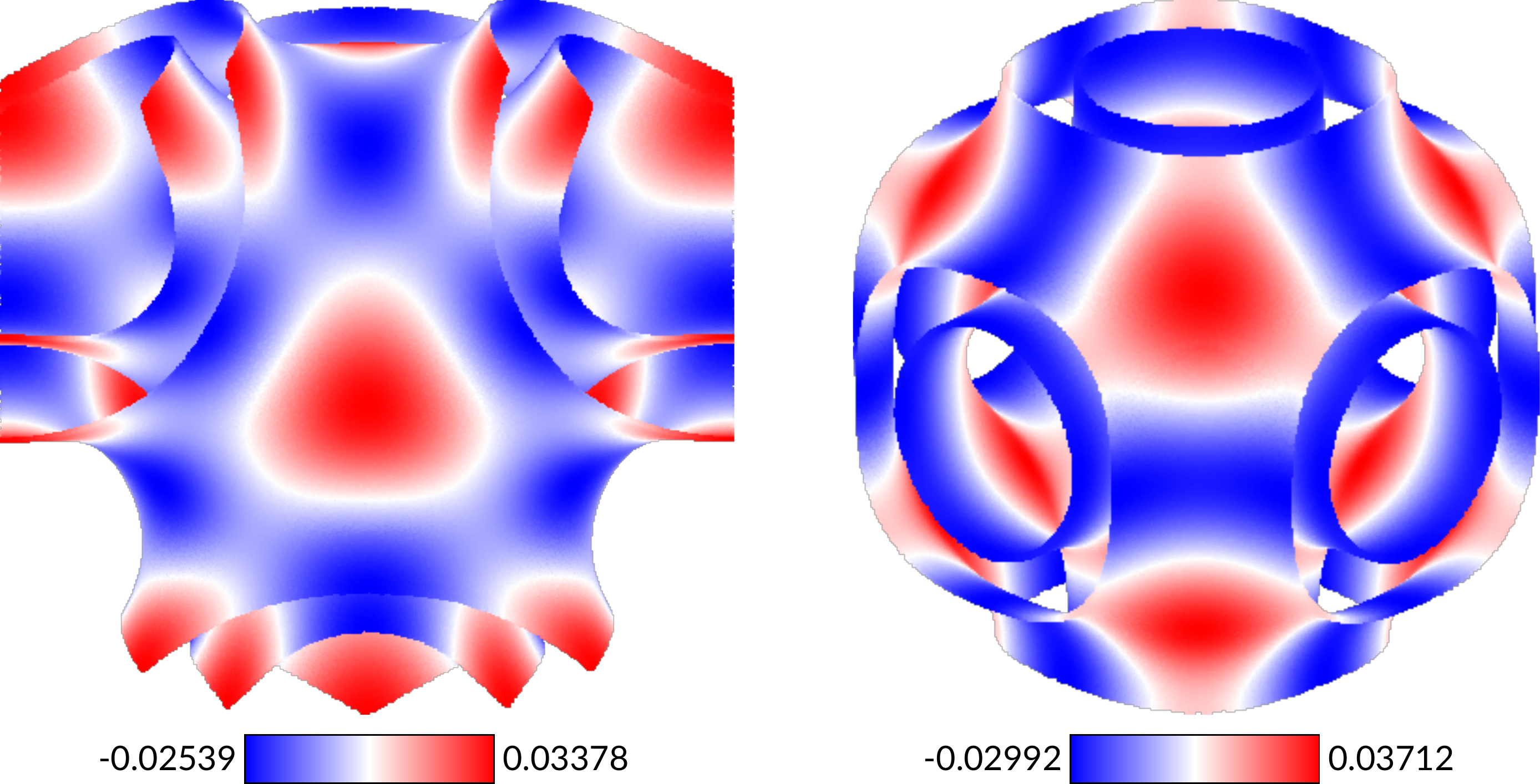}
  \caption{
  	Evolved neutral interfaces in the normal D (left) and P (right) phases with
  	$\lambda=10^4$, and the space between the neutral surfaces, which contains
  	the water, has a volume fraction $\phi_n=0.3$.  The color shows the
  	difference from the CMC surface with the same $\phi_n$: red indicates a
  	deviation away from the TPMS, and blue indicates a deviation towards the
  	TPMS.  The color bar below each figure shows maximum deviations, normalized
  	with unit lattice parameter.  Clearly, the competition causes expansions in
  	the necks.\label{fig:DiffNorm}
	}
\end{figure}

Plots of average chain length and average mean curvature are overlapped in
Figure~\ref{fig:meanHL-nor}, showing the geometric similarity between CMC model
and our model.  Evolved D and P bilayers with $\phi_n=0.3$ are shown in
Figure~\ref{fig:DiffNorm}, to compare with Figure~\ref{fig:DiffInv}.  Again,
the neutral interfaces tend to move away from the TPMS around the flat points,
and towards the TPMS at other places.  Indeed, the flat points of the TPMS are
most distant from the medial surface~\cite{schroder2003medial}.  As a
consequence, the pinching off necks gets expanded.  Comparing to the inverse
phases, the deviation is much larger ($>3\%$ of the lattice parameter).

\section{Conclusion}

We have demonstrated that modeling the competition between the curvature energy
and the chain stretching energy in LLC systems is possible with Surface
Evolver.  More specifically, we applied Willmore energy to triangulated
surfaces modeling the neutral interfaces, and Hooke energy to edges modeling
the hydrophobic chains.  A weighted sum of the two is used as the total
frustration of the system, and is minimized by Surface Evolver.  The resulting
surfaces are neither parallel surfaces nor CMC surfaces, but a compromise of
the two.  A detailed procedure is described and tested for LLCs in bicontinuous
cubic phases.

We compare our results on inverse cubic LLC phases to similar calculations with
the CMC model~\cite{shearman2007calculations}.  The squared variance of chain
length, as well as its contribution of the total frustration, is reduced as the
modulus of chain stretching energy increases.  This effect is more significant
in the P and D phases than in the G phase.  The frustration of the inverse G
phase eventually exceeds the inverse D and P phases.  A closer look reveals
that the pinching off necks in the CMC model, which was the main obstacle to
achieve a high dehydration~\cite{shearman2010towards, templer1998gaussian}, get
expanded in our model.  These observations prove the validity and usefulness of
our model.  When real data is applied, our calculation yields a G--D--P phase
sequence with increasing water fraction, which has been observed in
experiments.  We also performed same calculations on normal cubic LLC phases.
Similar observations are made, but the deviation from the CMC model is much
more significant than in the inverse phases.

\section{Technical details}\label{sec:tech}

In the opinion of the authors, the modelling capacity of Surface Evolver is
underestimated.  One purpose of this paper is to present a non-trivial
application of Surface Evolver in the study of interfaces.  In this section,
we provide details of our calculation on bicontinuous LLC cubic phases, as a
reference for readers who are interested in carrying out similar computations.

\subsection{Surface preparation}

The cubical periodic minimal surfaces considered in this paper, namely Schwarz'
G, D and P surfaces, can be generated from a small ``fundamental patch''
(Fl\"achenst\"uck) by Euclidean motions like translations, reflections,
rotations and rotoreflections.  However, only translations and reflections
apply to the accompanied neutral interfaces that are of physical interests.
Reflectional symmetries are related to free boundary conditions, and
translational symmetries to periodic boundary conditions.  Rotational
symmetries are often related to fixed boundary conditions, but the neutral
interfaces do not subject to any fixed boundary condition.  Hence we only
consider the fundamental unit of the translational and reflectional symmetries.

For the P and D surfaces, the \texttt{.fe} datafiles are available on Brakke's
website
(\url{http://facstaff.susqu.edu/brakke/evolver/examples/periodic/periodic.html}).
The P patch in the datafile is directly usable.  We have to apply a rotation to
the D patch to obtain the desired unit.  For the G surface, we use a
$96$-facets datafile kindly provided by Gro{\ss}e-Brauckmann;
see~\cite{kgb1997}.  The G surface has no reflectional symmetry, hence the
fundamental unit is a translational unit cell.  In Surface Evolver, the
\texttt{torus} model is used for the G surface to impose periodic boundary
condition.

After loading the datafile into Surface Evolver, we use the command \texttt{r}
(refine) to subdivide each triangle into four and, after each refinement,
evolve the surface towards the minimum of the area functional.  A translational
unit cell is not a minimizer of the area functional, unless a volume constraint
is imposed~\cite{kgb1996}, which we did for the G surface.  Apart from the
command \texttt{g} that performs one iteration of gradient descent, the command
\texttt{hessian\_seek} applies Newton--Raphson method to the gradient of
energy.  It is more efficient, yet safer than the more radical \texttt{hessian}
command.

As in~\cite{shearman2010towards}, the P patch is refined four times, yielding
$1024$ faces per patch; and the D patch is refined five times, yielding $2048$
faces per patch.  Due to lack of CPU power, we only refine the G surface three
times, yielding $6144$ faces per unit cell\footnote{Shearman et al.\ also claim
to have refined three times in~\cite{shearman2007calculations}, but report
$24576$ facets.  Their program provided as supplemental material shows four
refinements.}.  Surface Evolver provides commands to eliminate elongated
triangles (\texttt{K}) and extreme edges (\texttt{l} and \texttt{t}), as well
as commands for vertex averaging (\texttt{V}) and equitriangulation
(\texttt{u}).  They are frequently employed to keep the triangulation in good
shape.

Surface Evolver is very good at preparing triangulations.  But if an exact
formula is known for the surface of interest, one could also use a mesh
generator to produce a triangulation of high quality; see for
instance~\cite{schroder2003medial}.

\subsection{Setting up the bilayer}

Now that we are in possession of a triangulation of high quality, we make two
copies of it to model the neutral interfaces.  The commands
\texttt{new\_vertex}, \texttt{new\_edge} and \texttt{new\_facet} are useful in
this step.  We also use element attributes to remember the correspondence
between the elements in the original triangulations and in the copies.

To model an inverse phase, we only need to create edges between the
corresponding vertices in the copied triangulations (model (a) as illustrated
on the left of Figure~\ref{fig:model}).  The edge length is thus twice the
length of the chain we intend to model.  This practice aims to eliminate the
interference of the original TPMS in our calculation.  After creation of the
chains, The TPMS is fixed, and serves only as a reference.  This procedure also
works on unbalanced TPMS, such as the I-WP surface.

To model a normal LLC phase, we need to find, for each vertex in the
triangulated TPMS, the nearest point on the medial surface along the normal
vector.  Efficient algorithms for this purpose based on Voronoi diagram was
proposed by Amenta et al.~\cite{amenta2001power} and Schr\"oder et
al.~\cite{schroder2003medial}.  In particular, \cite{schroder2003medial}
contains an extremely detailed description of the medial surfaces of the D, P
and G surfaces, accompanied by fantastic figures.  The medial surface of the P
surface is very simple: they are contained in the reflection planes.  Hence we
simply attach one end of the edge to a vertex $v$ on the P surface, extend the
edge along the normal vector at $v$ (available in the \texttt{vertex\_normal}
attribute), and attach the other end to the first point we encounter on the
reflection plane.  The same practice also yields a good approximation for the
medial surface of the D surface.

For the G surface, we implement with Scipy a slightly simplified version of the
algorithm in~\cite{amenta2001power}.  For a vertex of the triangulated G
surface, we choose from its Voronoi cell the vertex with the largest inner
product with its normal vector.  This is a good approximation of the ``pole''
in~\cite{amenta2001power}, which is defined as the furthest vertex of the
Voronoi cell.  Scipy computes Voronoi diagrams using the Qhull
library~\cite{qhull}.

Setting up for the inverse phases is automated with the script language of
Surface Evolver.  Automation for the normal phases is possible with exterior
programs.  

\subsection{Minimization and measurement}

In Surface Evolver, The Hooke energy $(\ell-\ell_0)^2$ is implmented as
\texttt{hooke\_energy}.   The quantity \texttt{star\_perp\_sq\_mean\_curvature}
is a robust implementation of the Willmore energy $(h-h_0)^2$;
see~\cite{hsu1992}.  The Willmore energy is internally weighted by the
effective area of each vertex, which is $1/3$ of the total area of facets
containing the vertex.  The parameters $h_0$ and $\ell_0$ are meant to be the
spontaneous mean curvature and the relaxed length.  We keep updating them to
the average values $\langle h \rangle$ and $\langle \ell \rangle$ so that the
energies measure the squared variances.  The surface tension energy is turned
off (\texttt{set facet tension 0}).

The moduli of the energies are also used for taking averages and
normalizations.  Let $\kappa_L$ and $\kappa_H$ be the experimental moduli for
the chain stretching and mean curvature energy, respectively.  Then in Surface
Evolver, the modulus for the chain stretching energy is $k_L a^2 A_v / a_0^2 A$
at vertex $v$, and the modulus for the curvature energy is $k_H a_0^2 / A a^2$
for the dimensionless squared variance.  Here, $A$ is the total area; $A_v$ is
the effective local area of the vertex $v$; $a$ is the experimental lattice
parammeter; $a_0$ is the lattice parameter in Surface Evolver model, which is
$2$ for the P and D surfaces, and $8$ for the G surface.  Recall that the
Willmore energy is internally weighted by effective area, hence a factor $1/A$
suffices for normalization.

Let $\phi_n$ be the desired fraction of the space between the neutral
interfaces.  We impose a volume constraint of value $\phi_n V$ to the body
bounded by the neutral interfaces.  Here $V$ is the volume of the fundamental
unit, which is $1/6$ for P, $2/3$ for D and $256$ for G surface.  For normal
phases, the neutral interfaces may intersect when $\phi_n$ is close to $0$.
This is physically impossible, but not a numerical mistake.  Surface Evolver
computes volume by vector integration on the bounding faces.  In our situation,
the integrations on the two neutral interfaces have opposite signs, and they
cancel to the imposed small volume.  As $\phi_n$ increases, the neutral
interfaces will be separated very soon.

In practice, the volume fraction $\phi_n$ is changed gradually in steps of
$0.01$.  The small changes not only ensure a precise measurement, but also
avoid brutal changes in the surface that could destroy the surface.  After each
step, the surface is evolved (by commands \texttt{g} and
\texttt{hessian\_seek}) towards the minimum of the total energy.  To measure
the dimensionless squared variances, the parameters $h_0$ and $\ell_0$ are
updated after each iteration to the area weighted average values $\langle h
\rangle$ and $\langle \ell \rangle$.  After the surface stabilizes, the values
of the energies are printed on screen by the command \texttt{Q}, or output to
an exterior file for future record.

To measure the chain stretching frustration in the CMC setting, we could simply
change the quantity \texttt{hooke\_energy} from \texttt{energy} to
\texttt{info\_only}.  Then the Hooke energy is calculated, but excluded from
the energy to minimize.  For the G phase, however, lack of free boundary
condition allows the surface to shift away in the absence of
\texttt{hooke\_energy}.  This would increase the variance of chain length.  To
avoid this, it is recommended to start with non-zero modulus of
\texttt{hooke\_energy}, change the modulus to $0$ after several iterations, and
change it back for measurements after obtaining a CMC.  In practice, we are
usually able to reduce the squared variance of mean curvature down to the order
of $10^{-26}$ or lower, hence the obtained surface is indeed of constant mean
curvature.

\subsection{Performance}

The program is very robust.  The author once accidentally applied the program
to a CMC gyroid, then the neutral interfaces evolved away from the original
position, and correctly stabilized beside the minimal gyroid.  In practice,
this means that one only needs a general idea of the surface to carry out our
method.  Meanwhile, brutal changes are still to be avoided: on higher
dimensional problems, gradient descent method often suffer from saddle points
and local minimums.  Hence we recommend incremental change of $\phi_n$ at small
steps.

Tilt of the chains are allowed, but is minimized by the \texttt{normal\_motion}
and \texttt{hessian\_normal} modes, which forces the vertices to move along the
normal directions of the surface.  This practice is very successful for inverse
phases, but does not work as well in normal phases.  With some more work, it is
possible to include the tilt into the energy to minimize.  This is however not
our current focus.

Distortion of the triangulation is the main source of inaccuracy.  Elongated
triangles and edges of extreme lengths would cause problems to Surface Evolver.
An initial triangulation of high quality is recommended for this reason. But
the problem is unavoidable when the bilayer becomes very thick.  In our
calculations, the measurement suffers from slowness and imprecision as $\phi_n$
approaches $0.50$.  Surface Evolver provides commands to modify the
triangulation and improve the quality.  But most of these commands involve
vertex deletion, which would destroy elastic edges.  Our options are limited to
two commands: \texttt{V} for averaging vertices, and \texttt{u} for
equitriangulation.

The measurement can be automated, but human supervision is recommended to
achieve a good precision.  The \texttt{dump} command saves intermediate status
of the surface, allowing the user to check for problems and perform additional
iterations to improve the precision.  A full measurement of G phase would take
a few hours on a personal laptop.

\bibliographystyle{alpha}
\bibliography{References}

\end{document}